\documentclass[twoside,twocolumn,9pt]{article}
\usepackage{extsizes}
\usepackage[super,sort&compress,comma]{natbib} 
\usepackage[version=3]{mhchem}
\usepackage[left=1.5cm, right=1.5cm, top=1.785cm, bottom=2.0cm]{geometry}
\usepackage{balance}
\usepackage{mathptmx}
\usepackage{sectsty}
\usepackage{graphicx} 
\usepackage{lastpage}
\usepackage[format=plain,justification=justified,singlelinecheck=false,font={stretch=1.125,small,sf},labelfont=bf,labelsep=space]{caption}
\usepackage{float}
\usepackage{fancyhdr}
\usepackage{fnpos}
\usepackage[english]{babel}
\addto{\captionsenglish}{%
  
}
\usepackage{array}
\usepackage{droidsans}
\usepackage{charter}
\usepackage[T1]{fontenc}
\usepackage[usenames,dvipsnames]{xcolor}
\usepackage{setspace}
\usepackage[compact]{titlesec}
\usepackage{hyperref}
\usepackage{epstopdf}%This line makes .eps figures into .pdf - please comment out if not required.
\usepackage{chemformula}
\usepackage[toc]{appendix}
\definecolor{cream}{RGB}{222,217,201}

\begin{document}

\pagestyle{fancy}
\thispagestyle{plain}
\fancypagestyle{plain}{
%%%HEADER%%%
\renewcommand{\headrulewidth}{0pt}
}
%%%END OF HEADER%%%

%%%PAGE SETUP - Please do not change any commands within this section%%%
\makeFNbottom
\makeatletter
\renewcommand\LARGE{\@setfontsize\LARGE{15pt}{17}}
\renewcommand\Large{\@setfontsize\Large{12pt}{14}}
\renewcommand\large{\@setfontsize\large{10pt}{12}}
\renewcommand\footnotesize{\@setfontsize\footnotesize{7pt}{10}}
\makeatother

\renewcommand{\thefootnote}{\fnsymbol{footnote}}
\renewcommand\footnoterule{\vspace*{1pt}% 
\color{cream}\hrule width 3.5in height 0.4pt \color{black}\vspace*{5pt}} 
\setcounter{secnumdepth}{5}

\makeatletter 
\renewcommand\@biblabel[1]{#1}            
\renewcommand\@makefntext[1]% 
{\noindent\makebox[0pt][r]{\@thefnmark\,}#1}
\makeatother 
\renewcommand{\figurename}{\small{Fig.}~}
\sectionfont{\sffamily\Large}
\subsectionfont{\normalsize}
\subsubsectionfont{\bf}
\setstretch{1.125} %In particular, please do not alter this line.
\setlength{\skip\footins}{0.8cm}
\setlength{\footnotesep}{0.25cm}
\setlength{\jot}{10pt}
\titlespacing*{\section}{0pt}{4pt}{4pt}
\titlespacing*{\subsection}{0pt}{15pt}{1pt}
%%%END OF PAGE SETUP%%%

%%%FOOTER%%%
\fancyfoot{}
\fancyfoot[LO,RE]{\vspace{-7.1pt}\includegraphics[height=9pt]{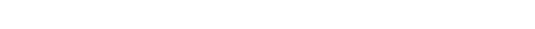}}
\fancyfoot[CO]{\vspace{-7.1pt}\hspace{13.2cm}\includegraphics{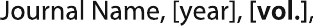}}
\fancyfoot[CE]{\vspace{-7.2pt}\hspace{-14.2cm}\includegraphics{head_foot/RF}}
\fancyfoot[RO]{\footnotesize{\sffamily{1--\pageref{LastPage} ~\textbar  \hspace{2pt}\thepage}}}
\fancyfoot[LE]{\footnotesize{\sffamily{\thepage~\textbar\hspace{3.45cm} 1--\pageref{LastPage}}}}
\fancyhead{}
\renewcommand{\headrulewidth}{0pt} 
\renewcommand{\footrulewidth}{0pt}
\setlength{\arrayrulewidth}{1pt}
\setlength{\columnsep}{6.5mm}
\setlength\bibsep{1pt}
%%%END OF FOOTER%%%

%%%FIGURE SETUP - please do not change any commands within this section%%%
\makeatletter 
\newlength{\figrulesep} 
\setlength{\figrulesep}{0.5\textfloatsep} 

\newcommand{\topfigrule}{\vspace*{-1pt}% 
\noindent{\color{cream}\rule[-\figrulesep]{\columnwidth}{1.5pt}} }

\newcommand{\botfigrule}{\vspace*{-2pt}% 
\noindent{\color{cream}\rule[\figrulesep]{\columnwidth}{1.5pt}} }

\newcommand{\dblfigrule}{\vspace*{-1pt}% 
\noindent{\color{cream}\rule[-\figrulesep]{\textwidth}{1.5pt}} }

\makeatother
%%%END OF FIGURE SETUP%%%

%%%TITLE, AUTHORS AND ABSTRACT%%%
\twocolumn[
  \begin{@twocolumnfalse}
{\includegraphics[height=30pt]{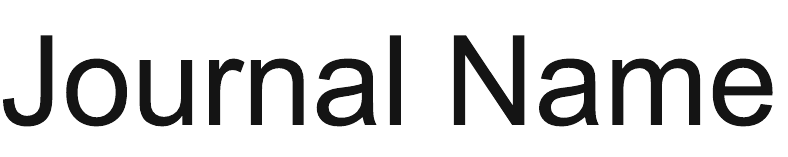}\hfill\raisebox{0pt}[0pt][0pt]{\includegraphics[height=55pt]{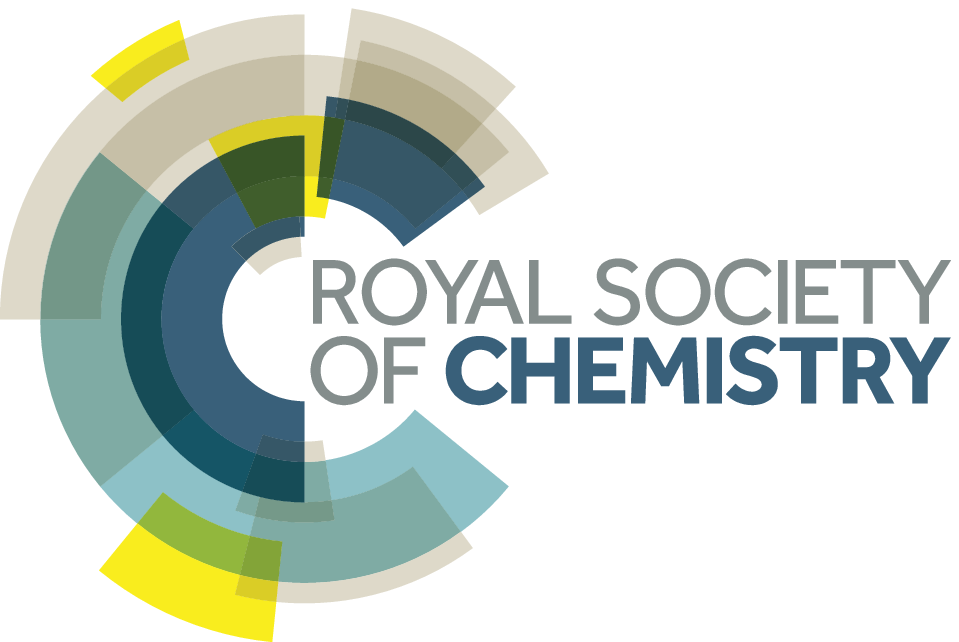}}\\[1ex]
\includegraphics[width=18.5cm]{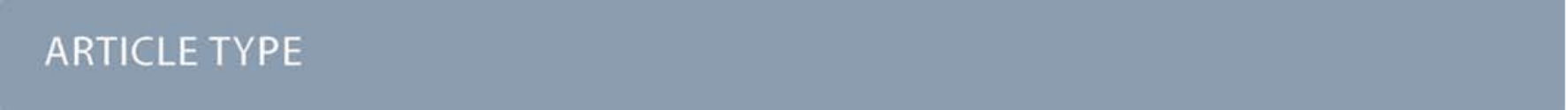}}\par
\vspace{1em}
\sffamily
\begin{tabular}{m{4.5cm} p{13.5cm} }

\includegraphics{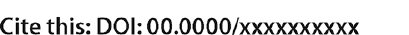} & \noindent\LARGE{\textbf{Iontronic microscopy of a tungsten microelectrode: “seeing” ionic currents under an optical microscope$^\dag$}} \\%Article title goes here instead of the text "This is the title"
\vspace{0.3cm} & \vspace{0.3cm} \\

 & \noindent\large{Zhu Zhang,$^{\ast}$\textit{$^{a}$} and Sanli Faez\textit{$^{a}$}} \\%Author names go here instead of "Full name", etc.

\includegraphics{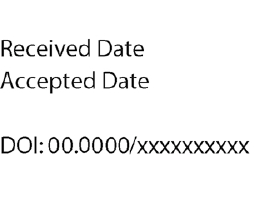} & \noindent\normalsize{Optical methods for monitoring the electrochemical reaction at the interface are advantageous because of their table-top setup and ease of integration into reactors. Here we apply EDL-modulation microscopy to one of the main components of amperometric measurement devices: a microelectrode. We present experimental measurements of the EDL-modulation contrast from the tip of a tungsten microelectrode at various electrochemical potentials inside a ferrocene-dimethanol \ch{Fe(MeOH)2} solution. By the combination of the dark-field scattering microscope and the lock-in detection technique, we measure the phase and amplitude of local ion-concentration oscillations in response to an AC potential as the electrode potential is scanned through the redox-activity window of the dissolved species. We present the amplitude and phase map of this response As such, this method can be used to study the spatial and temporal variations of the ion-flux due to an electrochemical reaction close to metallic and semiconducting objects of general geometry. We discuss the advantages and possible extensions of using this microscopy method for wide-field imaging of ionic currents.} 

\end{tabular}

 \end{@twocolumnfalse} \vspace{0.6cm}

]
%%%END OF TITLE, AUTHORS AND ABSTRACT%%%

%%%FONT SETUP - please do not change any commands within this section
\renewcommand*\rmdefault{bch}\normalfont\upshape
\rmfamily
\section*{}
\vspace{-1cm}

%%%FOOTNOTES%%%

\footnotetext{\textit{$^{a}$~Nanophotonics, Debye Institute for Nanomaterials Science, Utrecht University, 3584CC Utrecht, The Netherlands; E-mail:  z.zhang@uu.nl}}

%Please use \dag to cite the ESI in the main text of the article.
%If you article does not have ESI please remove the the \dag symbol from the title and the footnotetext below.
% \footnotetext{\dag~Electronic Supplementary Information (ESI) available: [details of any supplementary information available should be included here]. See DOI: 00.0000/00000000.}
%additional addresses can be cited as above using the lower-case letters, c, d, e... If all authors are from the same address, no letter is required

% \footnotetext{\ddag~Additional footnotes to the title and authors can be included \textit{e.g.}\ `Present address:' or `These authors contributed equally to this work' as above using the symbols: \ddag, \textsection, and \P. Please place the appropriate symbol next to the author's name and include a \texttt{\textbackslash footnotetext} entry in the the correct place in the list.}

%%%END OF FOOTNOTES%%%

%%%MAIN TEXT%%%%

\section{Introduction}
Electro-reflectance, electric field modulation of optical reflectivity, is a well-established method of investigating the nanoscale surface layer at metal-electrolyte interfaces ~\cite{feinleibElectroreflectanceMetals1966, prostakElectroreflectanceMetals1967, 
EffectElectricalDouble1968,
buckmanElectroreflectanceChangesDielectric1968,
parsonsBandStructureAssociated1969, bewickStudiesCathodicAdsorption1970}.
This signal is both sensitive to the carrier concentration on the metallic side as well as the ion concentration in the electric double layer~\cite{McIntyreElectrochemicalModulationSpectroscopy1973}. 
It has been used in the past to investigate band structures of the conducting side, as well as the electrochemical processes at the interface. 
The early measurements of electroreflectance were using the far-field reflection from flat surfaces, similar to ellipsometry, to measure the minuscule changes of the reflectivity due to the restructuring of the interfacial layer.
In those geometries, the spatial resolution is given by the illumination beam size, which is not suitable to investigate sub-ten-micrometer spatial variations.
Obtaining real-space surface images of electrochemical activity using optical reflection was made possible by using interferometric methods~\cite{liInterferometricMeasurementDepletion1995a, flatgenTwoDimensionalImagingPotential1995, anderssonImagingSPRDetection2008} or using surface plasmon imaging~\cite{shanImagingLocalElectrochemical2010}.
Plasmonic-enhanced measurements, however, are mainly suitable for the surface of noble metals such as gold.

Adapting and applying this powerful optical method to the nanoscale non-plasmonic structures and interfaces has proven difficult because of the optical diffraction limit and steep decrease of the scattering signal from sub-diffraction nano-objects. 
Metallic nanoparticles that exhibit a plasmonic resonance~\cite{kirchnerSnapshotHyperspectralImaging2018, byersSingleParticleSpectroscopyReveals2014, byersSingleParticlePlasmonVoltammetry2016, hoenerSpectralResponsePlasmonic2017, hoenerSpectroelectrochemistryHalideAnion2016, byersTunableCoreshellNanoparticles2015, hoenerPlasmonicSensingControl2018} and some 2D materials~\cite{zhuOpticalImagingCharges2019a},  have been notable exceptions. 
In this domain, the resonance enhancement of scattering allows, for example, the investigation of nanorods as small as 20 nanometers~\cite{hoenerSpectroelectrochemistryHalideAnion2016, hoenerSpectralResponsePlasmonic2017}. 
The signal in those investigations, however, is dominated by the electronic density inside the particle, and not all metallic particles exhibit a clear plasmon resonance in the visible range. 
Furthermore, as the career dynamics in the metallic side of the interface dominate the signal, investigation of the ionic currents in the electrolyte will only be possible via indirect interpretations and modeling.

Recently, our group has demonstrated a new optical contrast mechanism for non-plasmonic or dielectric particles based on a periodic modulation of the substrate potential~\cite{naminkElectricDoubleLayerModulationMicroscopy2020} relative to an electrode in the bulk of the liquid. 
In this method, modulating the EDL close to the surface results in scattering signals that are sensitive to both the local topography and electrochemical properties of the investigated region. 
Interestingly, the relative contrast in this interferometrically enhanced method increases with decreasing size of the particle. 
We have dubbed our method EDL-modulation microscopy, which can be categorized as a subclass of iontronic microscopy methods.
Recent advances in interferometric scattering microscopy and computational modeling of the electric double layer formation indicate that sensing a single surface charge alteration is within technical reach~\cite{zhangComputingLocalIon2021}.
Merryweather \textit{et al.} resolved nanoscopic lithium-ion dynamics in solid-state battery based on interferometric scattering microscopy\cite{merryweatherOperandoOpticalTracking2021, merryweatherOperandoMonitoringSingleparticle2022}.
Valavanis \textit{et al.} demonstrated the combination of scanning electrochemical cell microscopy (SECCM) and interference reflection microscopy (IRM) to monitor interfacial processes the SECCM meniscus status with high spatial and temporal resolution\cite{valavanisHybridScanningElectrochemical2022}.
Utterback \textit{et al.} spatiotemporally  resolved  electrochemically-induced ion concentration gradient evolution in solution using the interference  reflection  microscopy\cite{utterbackOperandoLabelfreeOptical2023}.

In this contribution, we present experimental measurements of the EDL-modulation contrast from the tip of a tungsten microelectrode at various electrochemical potentials inside a Ferrocene-dimethanol (\ch{Fe(MeOH)2}) solution. 
This is possible as a unique feature of EDL-modulation microscopy because the observed signal is not dependent on local resonances.
Our main goal is to investigate the spatial extent of the electrochemical reaction zone close to the microelectrode tip, with optical imaging.
Contrary to our previous measurements on transparent ITO substrates, measuring the microelectrode allows us to make a direct correlation between the electric current passing through the tip of the electrode and the observed optical modulation signal. 
While, in this paper, we use a confocal scanning method for building the image, this microscopy method is in principle compatible with wide-field imaging on a camera.
Achieving a simultaneously high temporal and high spatial resolution is essential for filtering out the electrochemical noise caused by surface heterogeneity, which is a nuisance in conventional amperometry measurements. 
Furthermore, this dynamic and correlative measurement allows us to separate the contribution of surface ion absorption to the potentiodynamic scattering signal from that of the diffuse double layer, which in turn can provide new insights into the dynamics of complex surface electrochemical processes.

In the following, we first describe our measurement setup and experimental conditions. 
Next, we present the signal of the lock-in enhanced EDL-modulation optical imaging from the tip of the microelectrode and discuss its dependence on the concentration of reagents, the modulation frequency and the modulation amplitude.
Finally, we present the confocal maps of the intensity and detected modulation phase around the tip.
We will conclude by presenting our perspective on using the EDL-modulation microscopy method for electrochemical imaging.

\section{Results}

\subsection{measurement setup}

The EDL-modulation microscopy setup is based on total-internal-reflection (TIR) illumination and scattering. The evanescent light field, incident on the interface, illuminates objects that are in the roughly 200~nm vicinity of the glass slide and inside the solution. 

\begin{figure}[h!]
    \centering
    \includegraphics[width=8cm]{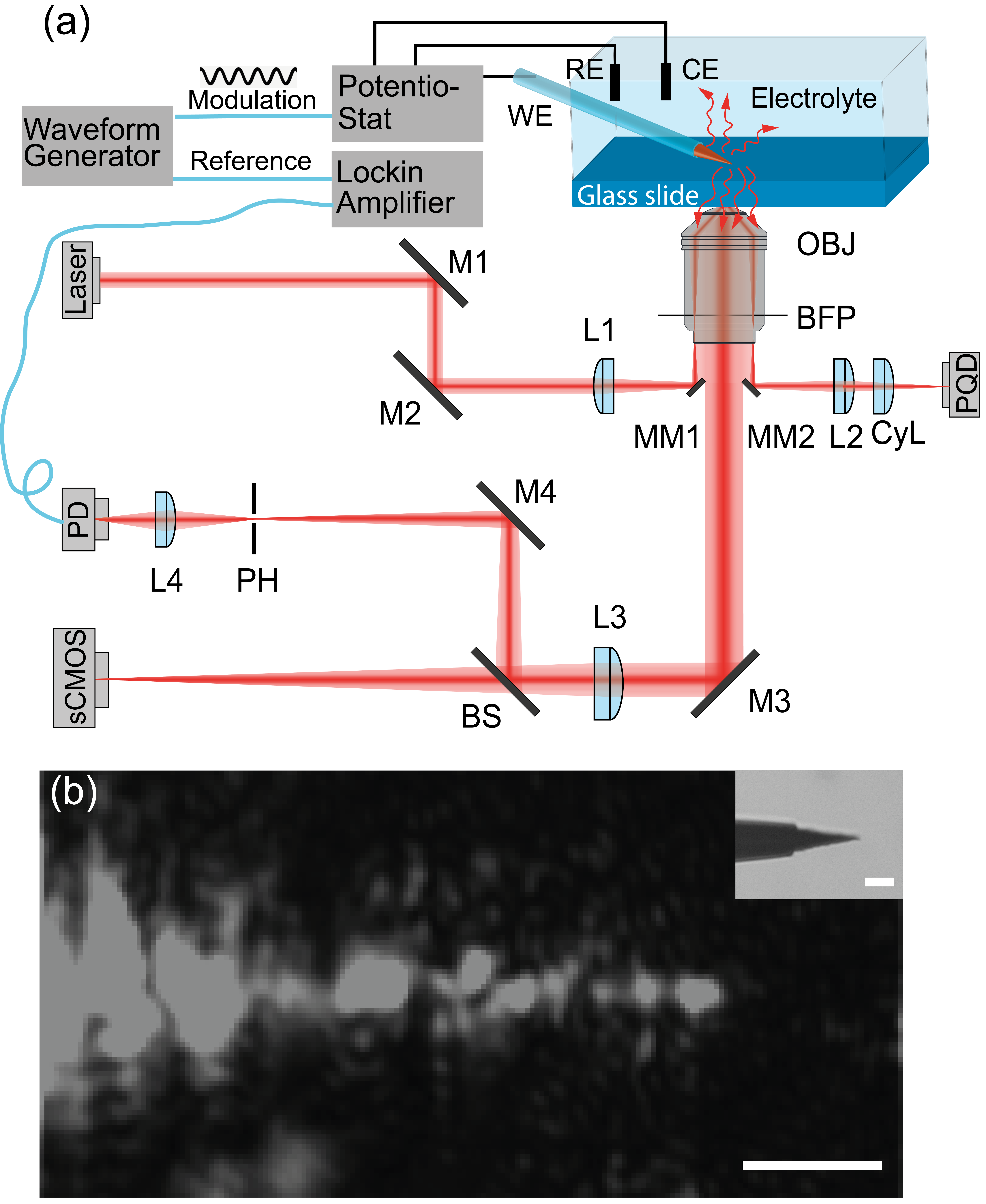}
    \caption{(a) Schematic of the setup. M1-M4: adjustable mirrors; L1,L2: beam focus lens; L3, L4: imaging lens; MM1, MM2: small prism mirrors; OBJ: microscope objective; BFP: back focal plane of the microscope objective; CyL: cylindrical lens; RE: reference electrode in the chemical cell; CE: the counter electrode in the chemical cell; BS: beam splitter; PH: pinhole; PD: photodiode; sCMOS: scientific complementary metal–oxide–semiconductor camera; QPD: quadrant photodiode. (b) Light-scattering image of the tungsten tip, inset: the bright-field image of the tungsten tip. Scale bars: 5 $\mu m$}
    \label{fig:setup}
\end{figure}

%% FIG1 setup
Fig.\ref{fig:setup} depicts the setup. A laser beam ( ignis, 640~nm, Laser quantum) is focused off-axis in the back focal plane of an oil-immersion objective (Nikon, CFI Apochromat TIRF $60\times$, 1.4~NA), giving rise to TIR.
The sensitive signal is derived from the scattering of the evanescent light field, which is imaged by a scientific complementary metal-oxide semiconductor (sCMOS) camera (ORCA-Flash 4.0 V3, Hamamatsu) and by a Photodetector (PD, PWPR-2K-SI, FEMOTO).

%%% end of fig environment

This method is similar to Total-Internal-Reflection-Fluorescence microscopy (TIRF), which excites the fluorescent objectives at the interface with an evanescent field and images its corresponding fluorescence emission light~\cite{axelrodCellsubstrateContactsIlluminated1981, thompsonMeasuringSurfaceDynamics1981, axelrodTotalInternalReflection2001}.
Instead of imaging fluorescence emission light from objectives, the experimental instrument detects the scattering light that reaches objectives.
That means it allows us to image any dielectric object that is closer than roughly 200~nm to the interface. 
Meng \textit{et al.} implemented this method to track single gold nanoparticles at oil-water interfaces~\cite{mengMicromirrorTotalInternal2021c} and Namink \textit{et al.} use this method to image electric double charging/discharging dynamics around small ITO nanoparticles during potential modulation~\cite{naminkElectricDoubleLayerModulationMicroscopy2020}.
For this paper, we have improved the detection sensitivity of our setup by including a photodiode and a lock-in amplifier in the imaging path, which improves the detection sensitivity, modulation divided by net scattering, from ~$10^{-3}$~\cite{naminkElectricDoubleLayerModulationMicroscopy2020, mengMicromirrorTotalInternal2021c} to ~$10^{-5}-10^{-6}$ using lock-in detection.
This extra sensitivity comes at the cost of imaging speed as one needs to scan the sample to build an image.

We use a 3D piezo stage (P-611.3 Nanocube, Physik Instrumente) to hold the glass slide and the microelectrode, and the piezo stage can precisely scan the sample in XY plane to get a 2D image.
The microelectrode we used in the experiments is a tungsten microelectrode (Microelectrodes Ltd., Cambridge, UK).
In the inset of Fig.\ref{fig:setup} (b), we show the bright field image of the microelectrode tip. 
The length of the exposed tip is about~15$\mu$m with a tip diameter of ~2$\mu$m at the very end, with a diameter of~10$\mu$m close to the glass insulation layer.
The rest of the microelectrode is insulated by a thin glass layer with a thickness of~1$\mu$m close to the tip and a thickness of~40$u$m at the shank.

We use a high-precision waveform generator (33120A, 15MHz, HP) to generate a modulation signal and a reference.
The modulated signal is sent to a potentiostat (E162 picostat, or EA362 Dual picostat, eDAQ), which accurately controls the cell potential during the experiments.
The reference signal is sent to the lock-in amplifier.  
The corresponding potential and current signal from the electrochemical cell are recorded and amplified by a data recorder (e-corder 410, eDAQ).
During the potential modulation, the variation in the scattering light intensity from the investigated objects is collected by the PD, which converts light to current and amplifies the current to voltage signal (with bandwidth 2K and transimpedance gain  of $10^{9} \Omega$). 
The amplified light intensity is sent to a lock-in amplifier (SR830 DSP lock-in Amplifier, Stanford Research System).
The lock-in amplifier calculates the amplitude of the light intensity variations caused by the potential modulation. 
We use a Data Acquisition card (DAQ, NIUSB-6212, National Instrument) to synchronize the waveform generator, lock-in amplifier, potentiostat and collect all the signals through a $Python$ program. 

It is possible to combine the lock-in method with wide-field imaging using an sCMOS camera and programming fast detection electronics~\cite{kurosuLabelFreeVisualizationNanoThick2022}. 
However, the discussion of such advanced imaging modalities is beyond the scope of this paper.  
To get a stable signal, the setup and illumination beam must be actively stabilized. 
These details are described elsewhere~\cite{Zhang2023}.

\subsection{The sample}

We perform scanning EDL-modulation measurement on a tungsten microelectrode (Microelectrodes Ltd., Cambridge, UK) inside \ch{Fe(MeOH)2}, dissolved in 100~mM KCl solution as support electrolyte.

\begin{figure}[h!]
    \centering
    \includegraphics[width=8cm]{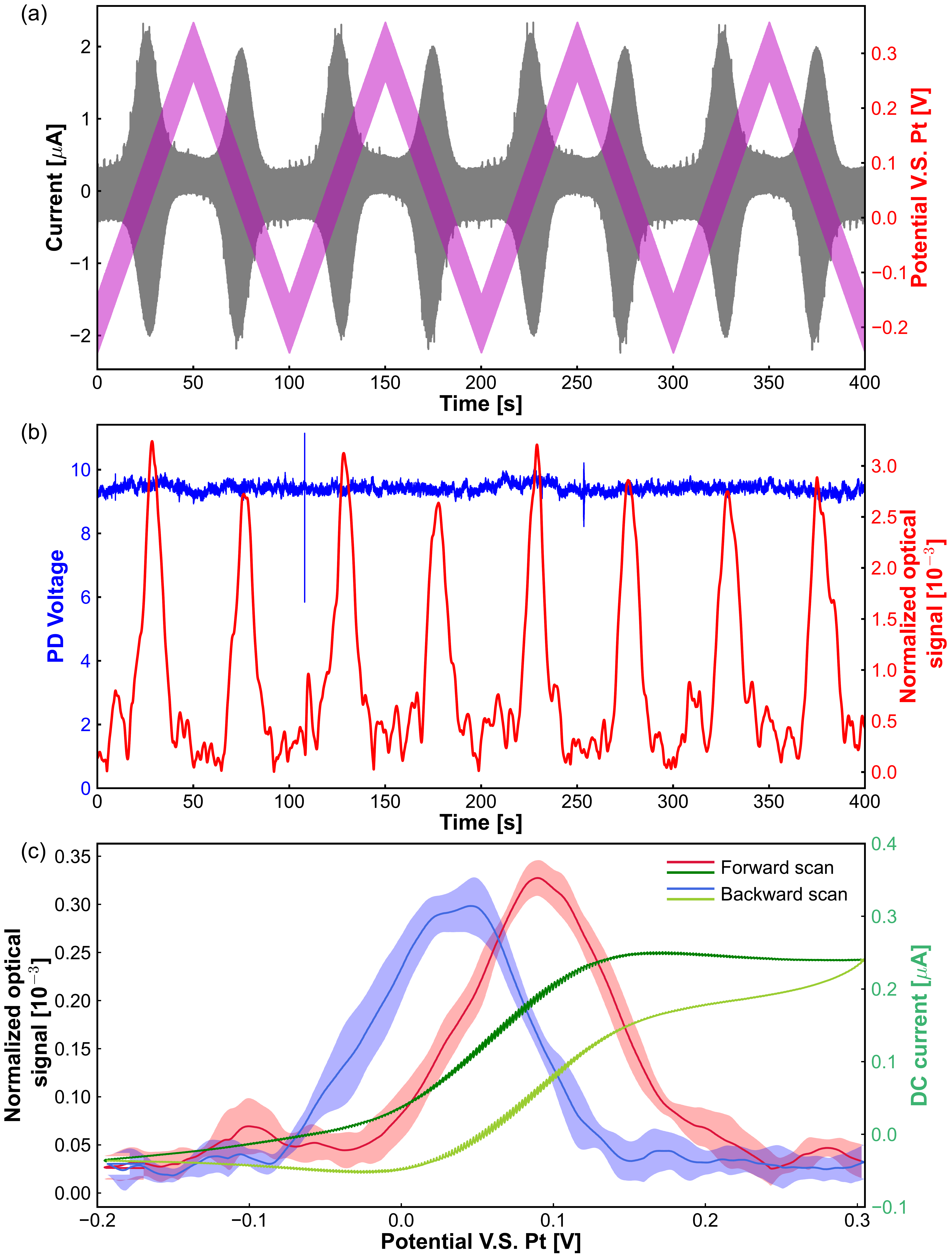}
    \caption{(a) Total ac current voltammogram (grey): the measured total electrical current through the microelectrode and the potential (pink) is scanned with a sine-modulated AC potential of 50~mV amplitude added to a gradually scanning DC offset. (b) Optical amplitude of lock-in signal (red) and the PD signal (blue). (c) Averaged optical amplitude of lock-in signal (blue and red) and the DC current component (green) extracted from total AC current voltammogram. DC potential scan from -200 mV to 300 mV with a scanning rate of 10~mV/s, frequency of potential modulation is 75~Hz, the amplitude of potential modulation is 50~mV, and the concentration of \ch{Fe(MeOH)2} is 5~mM.}
    \label{fig:timescane}
\end{figure}

In the experiment, we add a potential modulation to the offset potential scanned with a triangular shape. In conventional amperometric measurements, this method is best known as AC-voltammetry (ACV).
In ACV measurements, we refer to the offset as DC potential scan, and the potential modulation as AC potential, their corresponding currents are DC currents and AC currents.

As shown in Fig.~\ref{fig:timescane}(a), the purple curve presents the potential applied to the tungsten electrode, which is the superposition of an AC sinusoidal modulation potential with amplitude of 50~mV, frequency of 75~Hz and a linear triangle offset potential altering from $-$0.25~V to 0.3~V. 
The currents from the redox reaction of~\ch{Fe(MeOH)2} are shown by a grey line, we can see that the oscillation of the current is boosted in the~\ch{Fe(MeOH)2} redox potential window (around the potential of 0.05~V). 
Because of the potential modulation to the tungsten electrode, the light scattering intensities oscillate with respect to the potential, and the oscillation amplitude of the optical signal is detected by the lock-in amplifier. 

The red curve in Fig~\ref{fig:timescane}(b) shows the oscillation amplitude of the optical signal as a function of the DC scan potential.
The blue curve in Fig~\ref{fig:timescane}(b) represents the light scattering intensities collected by PD during the ACV measurement.
It is difficult to distinguish the changes or any trend by eye from the PD signal. However, with the lock-in amplifier, we can get the corresponding oscillation amplitude of the optical signal from the PD signal.

In the potential window of the electrochemical reactions, the oscillation amplitude of the optical signal goes up and returns back to the base value(around $0.05\times 10^{-3}$).
Fig.~\ref{fig:timescane}(c) shows the optical signal as a function of applied DC potential, which is an average of 4 cycles from the red curve in Fig.~\ref{fig:timescane}(b)
The green and yellow lines are the (average) DC part extracted from the measured current passing through the microelectrode in Fig.~\ref{fig:timescane}(a). 

\begin{figure}[h!]
    \centering
    \includegraphics[width=8cm]{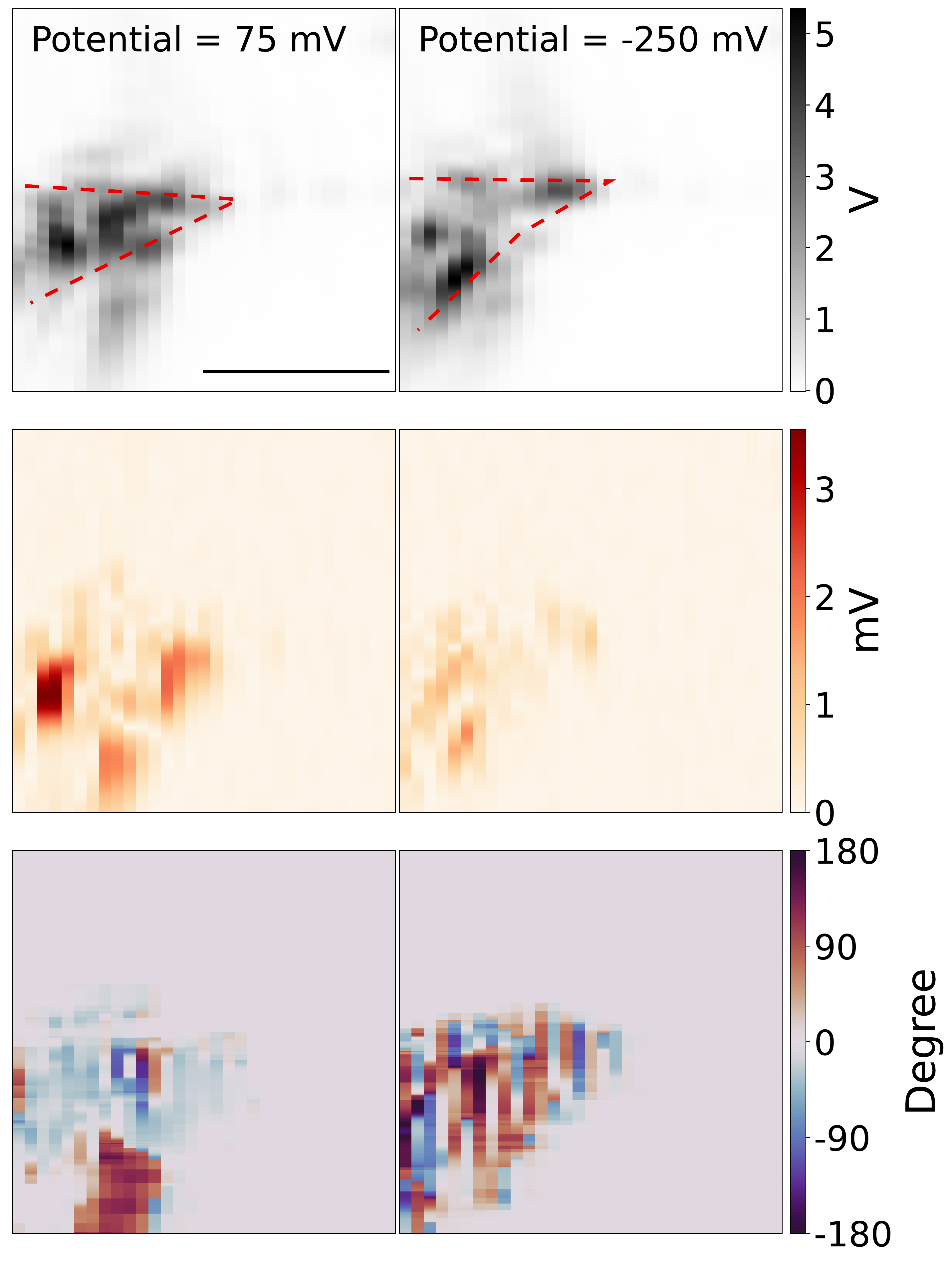}
    \caption{2D scan plot of the tip at potential 75~mV and $-$250~mV. Top panel: light intensity from Photodiode. middle panel: optical signal from lock-in amplifier. Bottom panel: phase plot. DC potential was kept at 75~mV (left) and $-$250~mV (right), in the potential modulation with the frequency of 95 Hz, and modulation amplitude of 50~mV, the concentration of \ch{Fe(MeOH)2} is 15~mM. Scale bar: 5~$\mu m$ }
    \label{fig:xyscan}
\end{figure}

Next, we use the piezo stage to scan the measurement spot. Fig.~\ref{fig:xyscan} (top panel) shows the 2D scan image of the tungsten tip, which is reconstructed from the scattering light intensity while the DC potential was kept at 75~mV and $-$250~mV, in the potential modulation with the frequency of 95~Hz and modulation amplitude of 50~mV. The red dashed line shows the rough shape and direction of the tip. 
Note that because of the TIR configuration, only parts of the tip that are in contact with the substrate can be visualized.
The middle panel shows the reconstructed 2D image from the optical lock-in signal.
The optical lock-in signal is higher around the tungsten tip at the redox potential window of  \ch{Fe(MeOH)2}  (left middle panel), while the potential is out of the redox potential window, the optical lock-in signal is weaker (right middle panel).

Note that the diffusion time of~\ch{Fe(MeOH)2^{0/+}},~\ch{K+} and ~\ch{Cl-} around the tungsten microelectrode tip is given by~$\tau_{i}$ = $r^{2}/D_{i}$, $r$ is the mean diameter of the exposed tip of tungsten microelectrode $r = (10+2)/2 = 6~\mu$m and $D_{i}$ denotes the diffusion coefficient of ~\ch{Fe(MeOH)2^{0/+}},~\ch{K+} and~\ch{Cl-}.  
With the diffusion constant  of $D_{Fc} = 6.3\times 10^{-10}$ m$^2$/s and $D_{\ch{K+}} \approx D_{\ch{Cl-}} \approx 2.0\times 10^{-9}$ m$^2$/s~\cite{vanmegenDifferentialCyclicVoltammetry2012, zevenbergenFastElectronTransferKinetics2009}, we can get the diffusion time $\tau_{Fc} = 1/70$~s and  $\tau_{\ch{K+}} \approx \tau_{\ch{Cl-}} \approx 1/200$~s.  In the 2D scanning experiments, the periodicity of the modulation (in few hundreds Hz) is shorter than the diffusion time $\tau_{i}$, which indicates the modulated diffusion of~\ch{Fe(MeOH)2^{0/+}} is  indeed around the tip area.
For the modulated diffusion of~\ch{K+} and~\ch{Cl-}, they can diffuse slightly further away from the tip than~\ch{Fe(MeOH)2^{0/+}}.

The tip shape changes due to the drift of the holder between these two measurements, which are taken about an hour separated from each other.
The bottom panel shows the phase of the lock-in response, which represents the delay between the local modulation of the scattering intensity and the applied potential to the tip.

Before discussing the two-dimensional scanning results, we discuss some control measurements obtained from the signal of the electrode tip.

\begin{figure}[h!]
    \centering
    \includegraphics[width=\columnwidth]{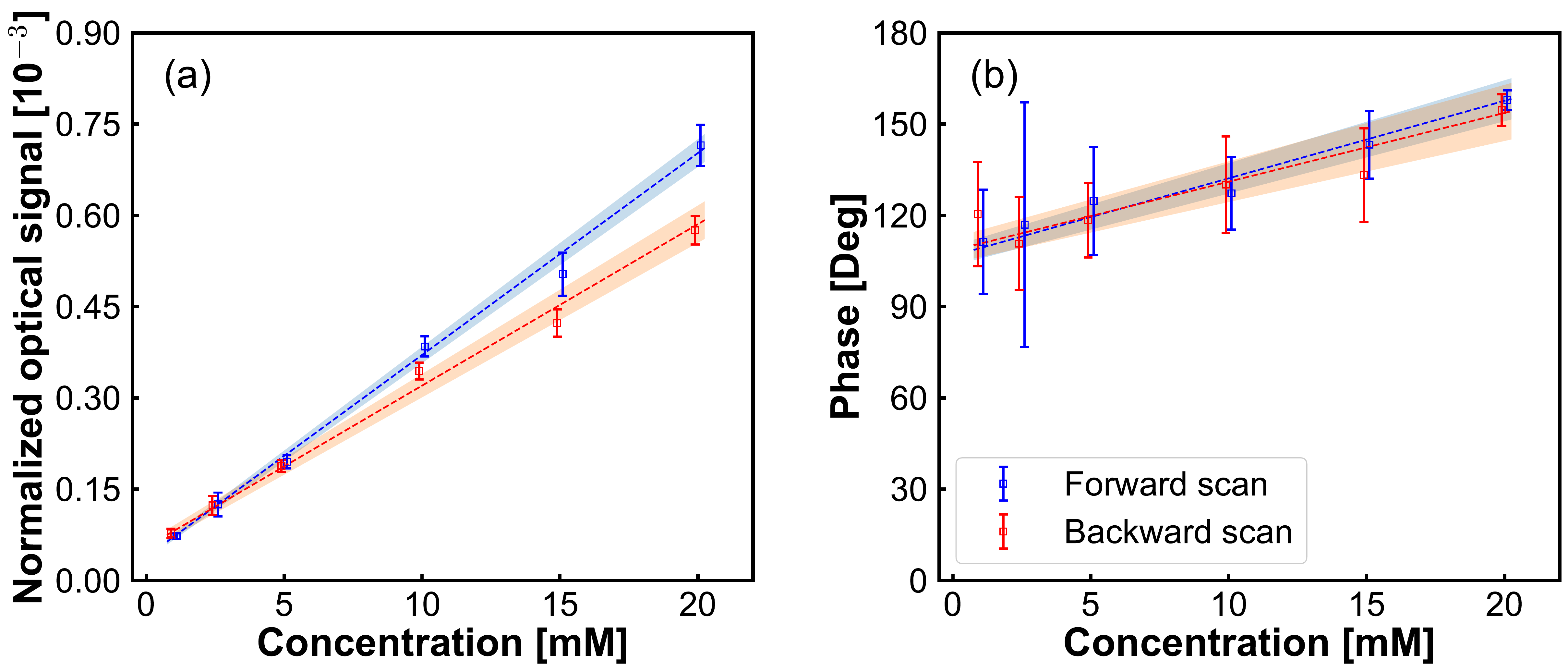}
    \caption{Normalized optical signal from lock-in amplifier with different \ch{Fe(MeOH)2} concentration. DC potential scan from $-$200 mV to 300 mV with a scanning rate of 40 mV/s, frequency of potential modulation is 75 Hz, the amplitude of potential modulation is 50 mV.}
    \label{fig:vsconcentration}
\end{figure}

In Fig.\ref{fig:vsconcentration} we have plotted the peak of the modulation signal, observed in the middle of the redox-reactivity window, as a function of~\ch{Fe(MeOH)2} concentration.
The current response ( containing AC component and DC component) as a function of~\ch{Fe(MeOH)2} concentration is provided in Appendix Figure \ref{fig:current_con}
An close to a linear relationship is observed, testifying to the sensitivity of the modulation signal to the concentration of the electrochemically active species.
Similarly, the measured signal scaled linearly with the modulation amplitude (Appendix Figure \ref{fig:vsamplitude}).
Both of these results are indications that the scattering modulation signal, for the conditions set in our experiments, is dominated by a first-order function of the analyte concentration in the solution adjacent to the electrode.

\begin{figure}[h!]
    \centering
    \includegraphics[width=\columnwidth]{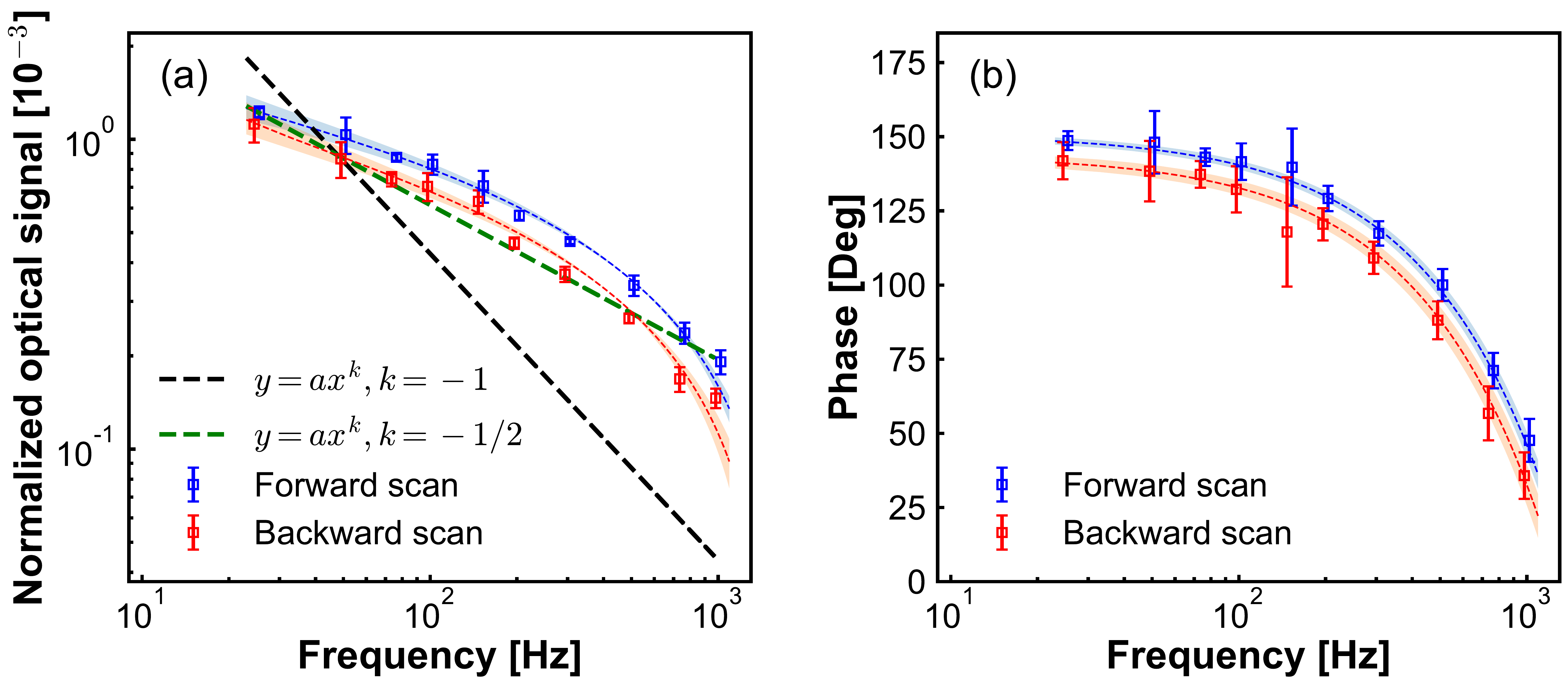}
    \caption{Normalized optical signal from lock-in amplifier with different frequency of potential modulation. DC potential scan from $-$200 mV to 300 mV with a scanning rate of 50 mV/s,  the amplitude of potential modulation is 50 mV, and the concentration of \ch{Fe(MeOH)2} is 15 mM. }
    \label{fig:vsfrequency}
\end{figure}

Next, we investigate the lock-in response dependence on the potential modulation frequency. 
These results are presented in Fig.\ref{fig:vsfrequency}. 
The lock-in signal decreases with increasing frequency, while the phase displays a semi-circle shape. 
Due to the limited bandwidth of our photodetector, we cannot be certain that the absolute phase lag at different frequencies is caused by the dynamics at the electrode tip and not by the response time of our photodetector when the frequency is higher than the bandwidth of our PD.  To avoid that, we keep the modulation frequency below the bandwidth (2kHz) of the PD. However, we still cannot exclude the effect of RC circuit response of the chemical cell on the amplitude and phase. 
We then have checked the electrical current response as a function of frequency modulation (Appendix Figure~\ref{fig:current_freq}). 
In Figure~\ref{fig:current_freq} we show the current response as a function of modulation frequency. 
Figure~\ref{fig:current_freq}(a) shows the amplitude of the AC current component, the slope of $1/2$ of the amplitude versus modulation frequency in the double logarithmic plot, which is the Randles-Sevick electron transfer process involving freely diffusing redox species. Figure~\ref{fig:current_freq}(b) shows the phase of the AC current component with respect to the phase of modulation potential, the dashed line is fitting to the data point to guide the eye. From the phase information of the AC current, we still cannot tell if the changes come from the modulation frequency difference or from the RC circuit.  Figure~\ref{fig:current_freq}(c) shows the DC current component, the DC current slowly increases with the increase of the modulation frequency.

In Fig.\ref{fig:vsfrequency}(a), the slope of $-1/2$ of modulation signal versus modulation frequency in the double logarithmic plot, resembles that for the Randles-Sevcik equation for cyclic voltammetry of electrochemically reversible electron transfer processes involving freely diffusing redox species ~\cite{elgrishiPracticalBeginnerGuide2018}. 
While the current increases proportional to the square root of the modulation frequency in conventional cyclic voltammetry (see Figure~\ref{fig:current_freq}(a)), here the measured optical signal decreases.
This difference can be understood by considering that the potentiodynamic signal in EDL-modulation microscopy originated from the ionic concentration variation at the electrode, as opposed to the ionic current variation in the conventional electrochemical measurement.

Following this analogy, if theoretically also verified, we can anticipate that our modulation frequency can also differentiate between the free diffusion effect on electrochemical processes and the influence of adsorbed species.

\begin{figure}[h!]
    \centering
    \includegraphics[width=\columnwidth]{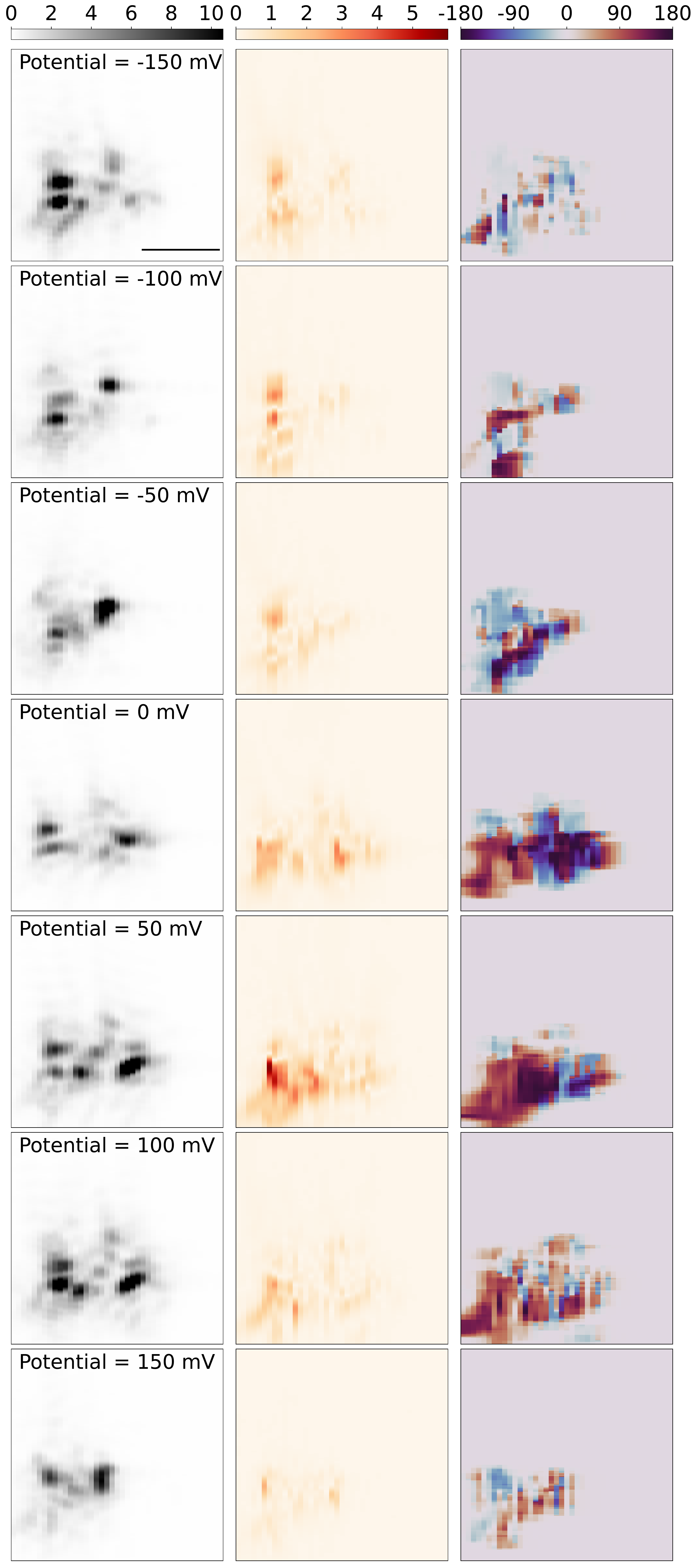}
    \caption{2D scan plot of the tip at different potentials.  Left panel: scattering light intensity from PD, unit: V. middle panel: optical signal from lock-in amplifier, unit: mV. Right panel: phase plot, unit: Degree. DC potential was kept at 0 mV (1st row), $-$100~mV (2nd row), $-$50~mV(3rd row) and 0~mV (4th row), 50~mV (5th row), 100~mV (6th row), 150~mV (7th row), in the potential modulation with the frequency of 95 Hz, and amplitude of 75~mV, the concentration of \ch{Fe(MeOH)2} was 15~mM. Scale bar: 5 $\mu m$ }
    \label{fig:xyscan_potential}
\end{figure}

Next, we present and discuss the spatial variations of the modulation amplitude, and most significantly its lock-in phase.
Fig.\ref{fig:xyscan_potential} displays the spatial map of the modulation signal around the tungsten tip as the sample stage is scanned over the objective.
The left panel shows the 2D-scan images of the tungsten tip, which is reconstructed from the scattering light intensity while the DC potential was kept at a constant potential both in and out of the redox potential window of \ch{Fe(MeOH)2},
in the potential modulation with the frequency of 95~Hz and modulation amplitude of 50~mV. 
The shape and direction of the tip are similar to the tip in Fig.~\ref{fig:xyscan}. 
Since the tip is barely in contact with the slide, during the measurements over hours, the tip can gradually drift on the slide surface, and hence change the scattering pattern slightly. 
The sliding behavior caused the contact angle and contract position changes can also result in discrepancies in the scattering light 2D images in different DC potentials. 
%This method is therefore recommended.

The middle panel in Fig.~\ref{fig:xyscan_potential} shows the reconstructed 2D images from the optical PD signal, amplitude and phase from lock-in response.
When the DC potential is kept out of the redox potential window of \ch{Fe(MeOH)2} -150~mV (1st row), and hence there is no electrochemical reaction around the tip, the optical lock-in signal shows the light scattering variations only induced by the \ch{K+} and \ch{Cl-} ions charging/discharging around the tungsten tip. 
While the DC potential is approaching from -150~mV to the redox potential of \ch{Fe(MeOH)2} at ~50mV, the oscillation amplitude of the optical signal increases. 
As the DC potential is closer to the redox potential, more \ch{Fe(MeOH)2} molecules are involved in the redox reactions.
With more \ch{Fe(MeOH)2} molecules redox reactions around the tip, the variation of the optical signal is larger. 
When the DC potential is kept in the redox potential window of \ch{Fe(MeOH)2} 50~mV (5th row), there are redox reactions of \ch{Fe(MeOH)2} and \ch{Fe(MeOH)2+}, 
as well as the \ch{K+} and \ch{Cl-} ions charging/discharging around the tip. 
The optical lock-in signal shows the light scattering variations both induced by the redox reaction of \ch{Fe(MeOH)2} and \ch{Fe(MeOH)2+}, as well as \ch{K+} and \ch{Cl-} ions charging/discharging around the tungsten tip. 

The right panel in Fig.~\ref{fig:xyscan_potential} shows the reconstructed 2D phase maps from the optical lock-in phase signal.
As the tungsten tip is scanning out of the view of the PD detection, the low input signal (mostly from background scattering) results in a small lock-in component and the identified phase value show sporadic variations. 
To avoid the random noise in the phase maps, the phase information is not displayed outside of the tip area, where the tip scattering light intensity is smaller than 10\% of the maximum intensity. 
As the frequency of the modulation is constant while scanning the measurement area over the tip, we can trust that the phase variation originates solely from the concentration dynamics of the redox-active species close to the electrode tip, and not the other electronic elements in the measurement circuit.

\begin{figure}[h!]
    \centering
    \includegraphics[width=\columnwidth]{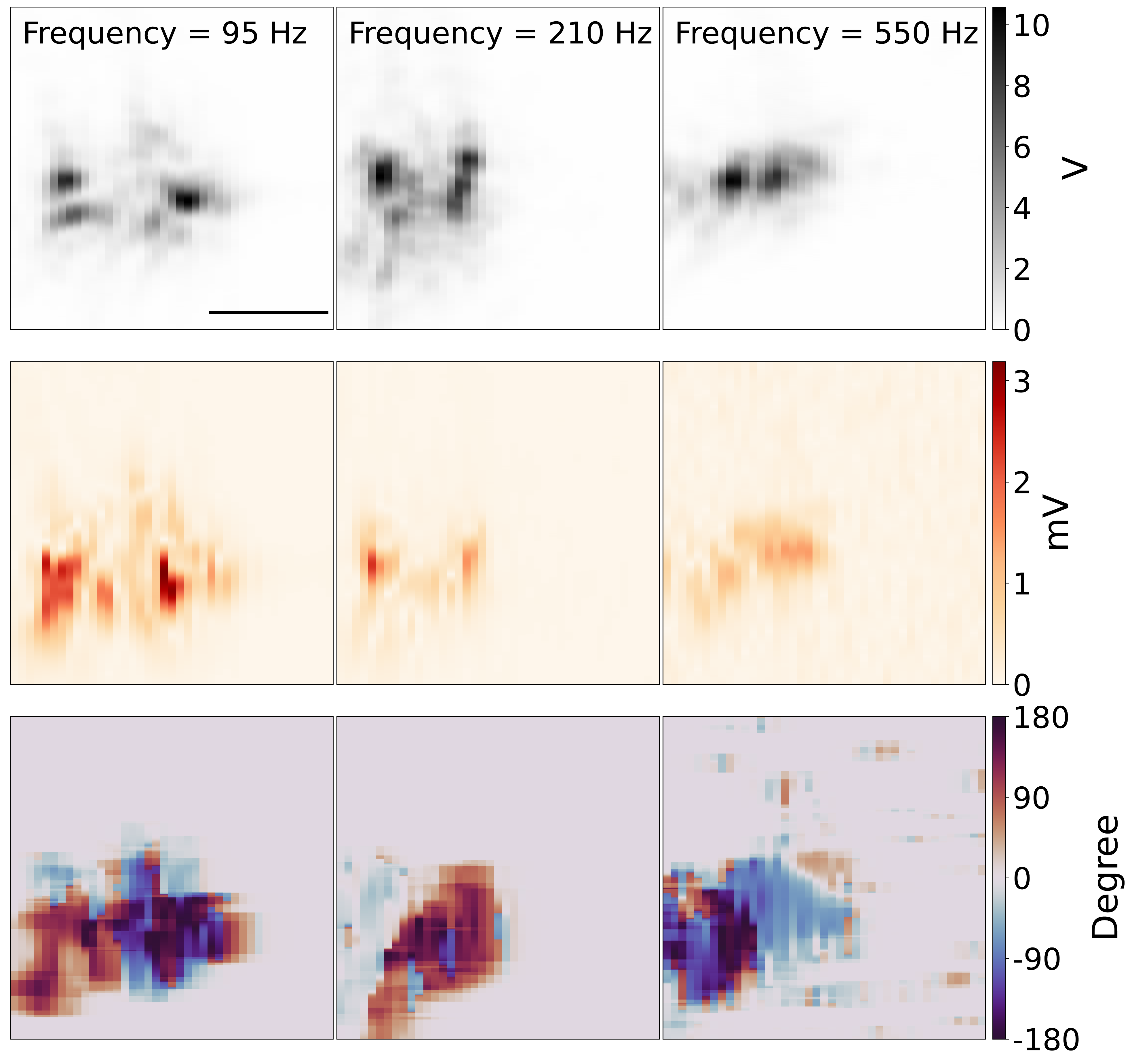}
    \caption{2D scan images of the tungsten tip at different modulation frequencies.  Top panel: light intensity from the photodiode. middle panel: optical signal from lock-in amplifier. Bottom panel: phase plot. AC modulation frequency was 95~Hz (1st column), 210~Hz (2nd column), 550~Hz (3rd column), DC potential was kept at 0~mV, in the potential modulation with the amplitude of 75~mV, the concentration of \ch{Fe(MeOH)2} was 15~mM. Scale bar: 5 $\mu m$. }
    \label{fig:xyscan_freqency}
\end{figure}

Fig.\ref{fig:xyscan_freqency} shows the 2D scanning imaging around the tungsten with different modulation frequencies. 
The top panel shows the 2D scan images of the tungsten tip, which is reconstructed from the scattering light intensity while the DC potential was kept at close to the redox potential window of \ch{Fe(MeOH)2} 50~mV, 
with the potential modulation amplitude of 75~mV. 
The shape and direction of the tip are similar to the tip shown in Fig.~\ref{fig:xyscan}. 

The middle row panel in Fig.\ref{fig:xyscan_freqency} shows the reconstructed 2D images from the optical PD signal, amplitude and phase of lock-in response.
Since the DC potential was kept around the redox potential of \ch{Fe(MeOH)2}, there are redox reactions of \ch{Fe(MeOH)2} and \ch{Fe(MeOH)2+}, 
as well as the \ch{K+} and \ch{Cl-} ions charging/discharging around the tungsten tip. 
The optical lock-in amplitude signal shows the light scattering variations, which is in this regime, linearly proportional to the ion concentration of the species with the highest optical polarisatbility. 
With lower modulation frequency, the optical lock-in signal is higher, and the signal area is larger. 
The redox species and \ch{K+} and \ch{Cl-} can diffuse further away from the tungsten.
With higher modulation frequency, the optical lock-in signal is weaker, and the 2D scanning signal area is smaller. 
The bottom row panel in Fig.\ref{fig:xyscan_freqency} shows the reconstructed 2D images from the optical lock-in phase signal.

As the DC potential and the amplitude of the modulation is constant while scanning the measurement area over the tungsten tip, 
with the higher modulation frequency, the lock-in signal is smaller, which results in a visual shrinking of the EDL-modulation image of the tip.
One interpretation of this observation can be that the density variation of the species and \ch{K+} and \ch{Cl-} can only diffuse to a limited distance away from the electrode surface.
The refractive index variations around the tungsten tip caused by the concentration of molecules and ions then is smaller compared to low frequency modulation.

\section{Discussions}

%some discussion points as bullets:

By combining the dark-field scattering microscope and the lock-in detection technique, we introduce a label-free and \textit{operando} method to measure the ion concentration variations during electrochemical reactions around a tungsten microelectrode with the equivalent of the AC-voltammetry technique. 
We have presented the alteration of this signal in the reversible electrochemical reaction window of \ch{Fe(MeOH)2}. 
Since the dark-field scattering microscope is based on the principle of total-internal-reflection,  the optical signal is only sensitive in the area where the tungsten tip contacts the glass slide, where it can scatter the evanescent field at the surface.

We have measured the optical signal with different AC modulation amplitude, AC modulation frequency, DC potential and different concentrations of \ch{Fe(MeOH)2}.
We found that the dynamics of ion  transport vary with both the AC modulation frequency and the concentration \ch{Fe(MeOH)2}.
The dynamics of ions transport also highly varies based on the surface topography and the contact between the electrode glass cover and the glass slide.

In this paper, we have shown a proof of principle of obtaining dynamic images with EDL-modulation microscopy. 
We can foresee that the method is suited more to measuring on structures fixed on the substrate, or by creating a pattern of nanoparticles as local transducers of the EDL-modulation signal.
To quantitatively analyze the dynamics of the ion transport, it is better to make a grid pattern of micro/nanoscopic objects on the surface, which can scatter light from the evanescent field close to the surface, such as nanoparticles array, nanodisk array\cite{kasaniReview2D3D2019} and nanowires array\cite{liuSelfassembledMagneticNanowire2007, suArtAligningOnedimensional2012}.
These well-defined structures also allow for stable measurements with longer time-scale and can therefore be used for systematic measurements with a range of different experimental parameters. 

Similar \textit{operando} images of ions transport inside micro/nanoparticles, such as the diffusion of lithium-ions in single microparticle, is observed by monitoring the relative change of scattering light intensity of the microparticle during lithiation and delithiation processes\cite{merryweatherOperandoOpticalTracking2021, merryweatherOperandoMonitoringSingleparticle2022}, the lithium-ions diffusion kinetics of single \ch{LiCoO2} nanoparticles are imaged by surface plasmon resonance microscopy (SPRM)\cite{jiangOpticalImagingPhase2017, sunCollisionOxidationSingle2017}, the electrochemical reaction kinetics of Prussian Blue (PB) Nanoparticles and \ch{K+} ions diffussion inside PBNPs are captered by SPRM\cite{jiangThinFilmElectrochemistrySingle2017}, total internal reflection microscopy\cite{yuanVerticalDiffusionIons2021} and dark-field microscopy\cite{niuDeterminingDepthSurface2022}, and the monitoring and differentiating of the electrodeposition dynamics of metallic Ni and \ch{Ni(OH)2} nanoparticles on ITO slide are introduced by imaging their bright-field optical contrast\cite{godeffroyDecipheringCompetitiveRoutes2021}.
The ions transport during electrochemical reactions in such environments or inside a solution can be a complex process following coupled nonlinear differential equations of motion, for each charged species, coupled to the Laplace equation for the electric field distribution~\cite{kragtRefractiveIndexMapping1990, liInterferometricMeasurementDepletion1995a, boonCoulombicSurfaceIonInteractions2023, utterbackOperandoLabelfreeOptical2023}. 

At this point, the dynamics of our system seem to be too complex to model with a quasi-stationary or one-dimensional approximation~\cite{lianBlessingCurseHow2020a, bohraModelingElectricalDouble2019, linMicroscopicModelCyclic2022} and one has to solve the complete time-dependent equations to resolve the dynamics of ion transport temporally and spatially~\cite{liuEnhancedElectrocatalyticCO22016, lemineurSituOpticalMonitoring2020, lemineurImagingQuantifyingFormation2021,
yangSimulatingChargingCylindrical2022, taoEnhancingElectrocatalyticN22022, utterbackOperandoLabelfreeOptical2023}.
However, the possibility of simultaneous measurements of the dynamic ion-density response around a micro- or nano-sctructure, as we have demonstrated in this paper, can be hugely beneficial for testing the validity of models proposed for the numerical simulation of such complex processes.
Meanwhile, using semi-heuristic models and calibration relative to well-studied processes, one can also use this technique as a labeling-free tool for electrochemical measurements in ultra-small volumes and for ultra-low current levels. 

\section*{Author Contributions}
%We strongly encourage authors to include author contributions and recommend using \href{https://casrai.org/credit/}{CRediT} for standardised contribution descriptions. Please refer to our general \href{https://www.rsc.org/journals-books-databases/journal-authors-reviewers/author-responsibilities/}{author guidelines} for more information about authorship.
Zhu Zhang:  Methodology, Experimental measurements,  Investigation, Visualization,  Writing Original draft preparation. Sanli Faez:  Conceptualization,  Methodology, Writing Original draft preparation, Writing, Reviewing and Editing, Supervision

\section*{Conflicts of interest}

There are no conflicts to declare.

\section*{Acknowledgements}
We thank Allard P. Mosk, Serge Lemay, Frédéric Kanoufi, Jean-François Lemineur, Louis Godeffroy and Haolan Tao for fruitful discussions. 
 We thank Paul Jurrius, Dante Killian, Aron Opheij and Dave J. van den Heuvel for technical support.
 We thank Microelectrodes Ltd. for providing the tungsten microelectrode.
This research was supported by the Netherlands Organization for Scientific Research (NWO grant 680.91.16.03), China  Scholarship  Council  (CSC  201806890015).

%%%END OF MAIN TEXT%%%


\balance


%%%REFERENCES%%%
\bibliography{FDIontronics_TunstenElectrode} %You need to replace "rsc" on this line with the name of your .bib file
\bibliographystyle{rsc} %the RSC's .bst file

%% Template instructions 

% For footnotes in the main text of the article please number the footnotes to avoid duplicate symbols. \textit{e.g.}\ \texttt{\textbackslash footnote[num]\{your text\}}. The corresponding author $\ast$ counts as footnote 1, ESI as footnote 2, \textit{e.g.}\ if there is no ESI, please start at [num]=[2], if ESI is cited in the title please start at [num]=[3] \textit{etc.} Please also cite the ESI within the main body of the text using \dag.

 \newpage
\onecolumn
\setcounter{figure}{0}
\renewcommand{\thefigure}{S\arabic{figure}}
\renewcommand{\figurename}{Figure}
\section*{Appendix}

\begin{figure}[h!]
    \centering
    \includegraphics[width=\columnwidth]{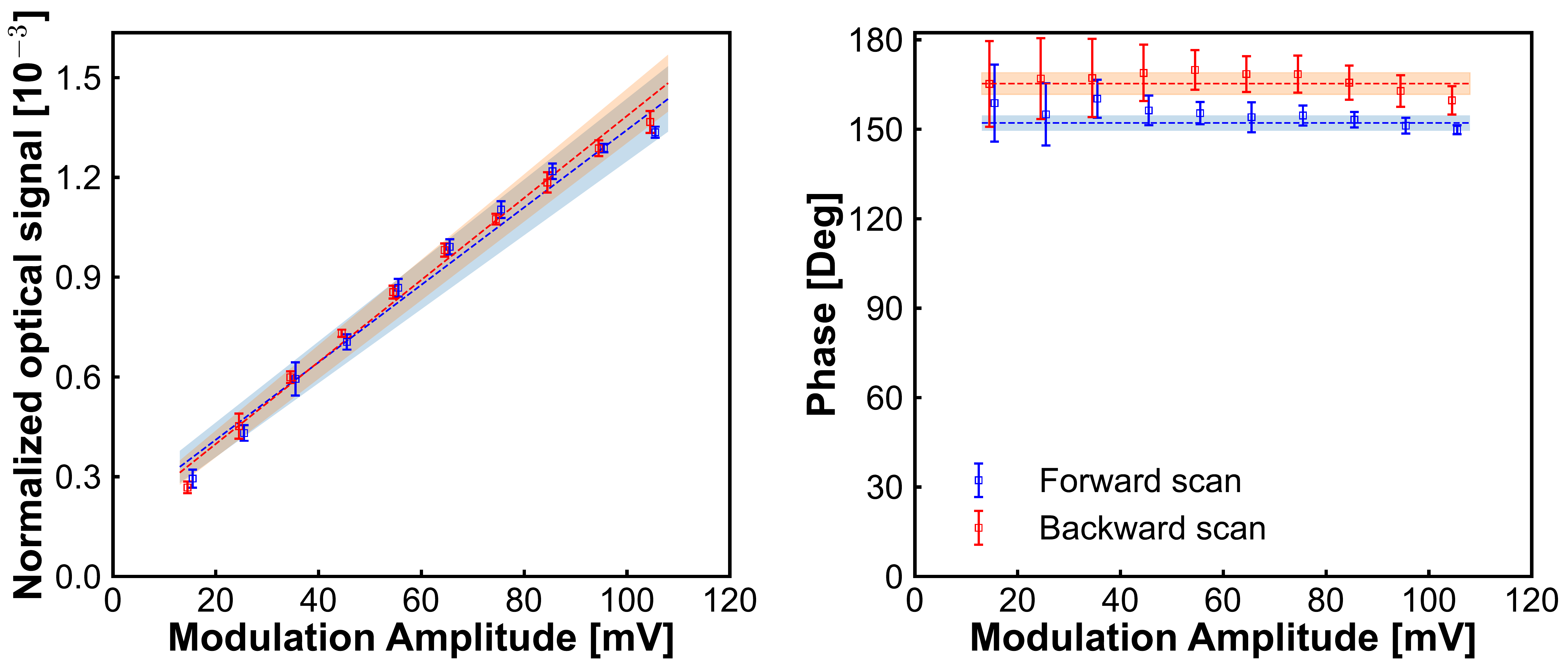}
    \caption{Normalized optical signal from lock-in amplifier with different modulation amplitude. DC potential scan from -200 mV to 300 mV with a scanning rate of 40 mV/s, frequency of potential modulation is 75 Hz, and the amplitude of potential modulation is 50 mV. At the concentration of 20~mM \ch{Fe(MeOH)2}.}
    \label{fig:vsamplitude}
\end{figure}

In Figure\ref{fig:vsamplitude} (a) we show the normalized optical amplitude of lock-in signal as a function of modulation amplitude. 
In Figure\ref{fig:vsamplitude} (b) we show the optical phase of lock-in signal as a function of modulation amplitude. 

\begin{figure}[h!]
    \centering
    \includegraphics[width=\columnwidth]{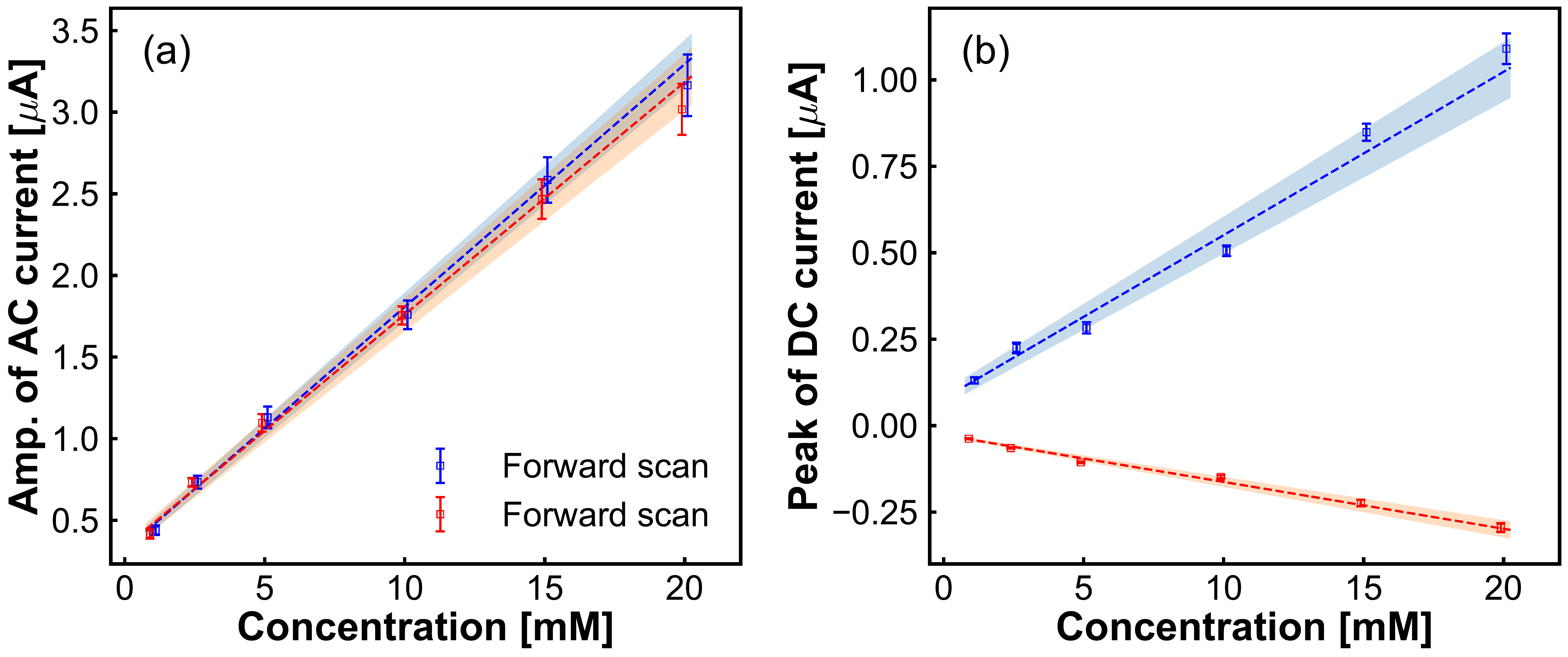}
    \caption{The amplitude of AC current component and DC current component as a function of~\ch{Fe(MeOH)2} concentration. (a) The amplitude of AC current component extracted from the measured total AC voltammogram. (b) The DC current component extracted from the measured total AC voltammogram. 
    DC potential scan from $-$200 mV to 300 mV with a scanning rate of 40 mV/s, frequency of potential modulation is 75~Hz, the amplitude of potential modulation is 50~mV.}
    \label{fig:current_con}
\end{figure}

In Figure\ref{fig:current_con} (a) we show the amplitude of AC current component extracted from the measured total AC voltammogram. 
To extract the AC current component, we first calculate the DC current component as shown in  in Figure\ref{fig:current_con} (b), by doing a moving averaging of the total AC voltammogram ( as shown in Fig.2 (a)) to remove the AC component.
Then we can get the AC current component by subtracting the calculated DC current component from the total AC voltammogram.

\begin{figure}[h!]
    \centering
    \includegraphics[width=\columnwidth]{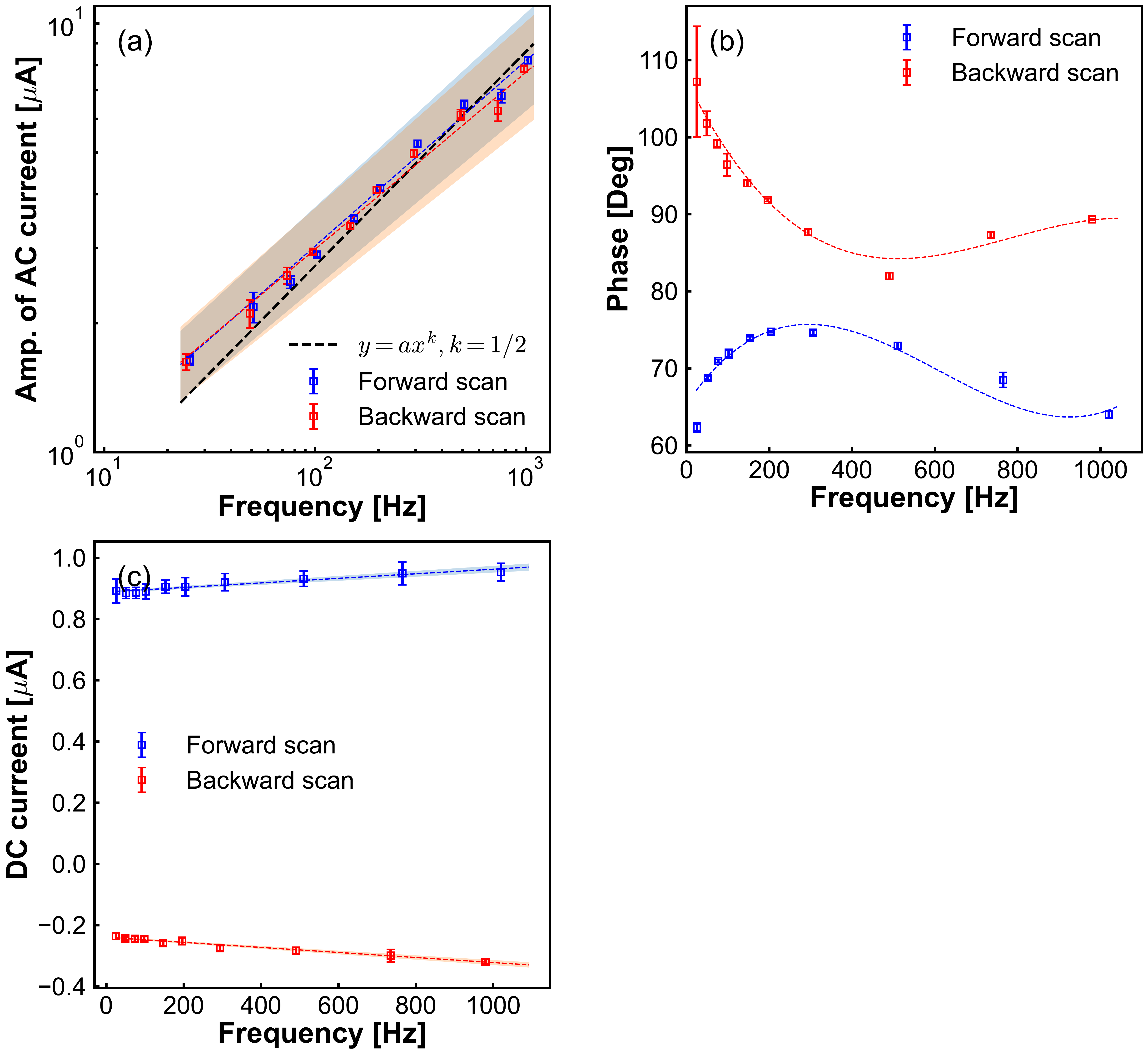}
    \caption{The current response as a function of modulation frequency. (a) The amplitude of AC current component as a function of modulation frequency. (b) The phase of AC current respect to the phase of modulation potential. The dashed is fitting to the data points to guide the eye.  (c) The peak of  DC current component.  DC potential scan from $-$200 mV to 300 mV with a scanning rate of 50 mV/s, the amplitude of potential modulation is 50 mV and the concentration of~\ch{Fe(MeOH)2} is 15 mM.}
    \label{fig:current_freq}
\end{figure}

In Figure\ref{fig:current_freq} we show the amplitude and phase of AC current component extracted from the measured total AC voltammogram, and the amplitude of DC current component. 

\end{document}